\documentclass[graybox]{svmult}

\usepackage{mathptmx}       
\usepackage{helvet}         
\usepackage{courier}        
\usepackage{type1cm}        

\usepackage{makeidx}         
\usepackage{graphicx}        
\usepackage{multicol}        
\usepackage[bottom]{footmisc}

\usepackage{amsfonts}
\usepackage{amssymb,amsmath}

\makeindex             

\def\ud{\mathrm{d}}

\begin{document}

\title*{Coherent states quantization and affine symmetry in quantum models of gravitational singularities}

\author{Herv\'{e} Bergeron \and Ewa Czuchry \and Przemys{\l}aw Ma{\l}kiewicz}


\institute{H. Bergeron \at ISMO, UMR 8214 CNRS, Univ Paris-Sud, France, \email{herve.bergeron@u-psud.fr} \and E. Czuchry \at National Centre for Nuclear Research, 00-681 Warszawa, Poland, \email{ewa.czuchry@ncbj.gov.pl}, \and P. Ma{\l}kiewicz \at National Centre for Nuclear Research,  00-681Warszawa, Poland, \email{Przemyslaw.Malkiewicz@ncbj.gov.pl}}

%
%
\maketitle


\abstract{We employ the framework of affine covariant quantization and associated semiclassical portrait to address two main issues in the domain of quantum gravitational systems: (i) the fate of singularities and (ii) the lack of external time. Our discussion is based on finite-dimensional, symmetry-reduced cosmological models. We show that the affine quantization of the cosmological dynamics removes the classical singularity and univocally establishes a unitary evolution. The semiclassical portrait based on the affine coherent states exhibits a big bounce replacing the big-bang singularity. \\
As a particularly interesting application, we derive and study a unitary quantum dynamics of the spatially homogenous, closed model, the Mixmaster universe. At the classical level it undergoes an infinite number of oscillations before collapsing into a big-crunch singularity. At the quantum level the singularity is shown to be replaced by adiabatic and nonadiabatic bounces. As another application, we consider the problem of time. We derive semiclassical portraits of quantum dynamics of the Friedman universe with respect to various internal degrees of freedom. Next we compare them and discuss the nature of quantum evolution of the gravitational field. }

\section{Introduction}
Affine covariant quantization pertains to a generic quantization framework based on operator-valued measures and named integral quantization \cite{hb2014, alijp2013, jpM2016, jpH2015, hb2017}. Integral Quantization (IQ) includes Coherent State (CS) quantization \cite{alijp2013, perelomov1986} and quantizations based on Lie groups, like the usual Weyl-Wigner quantization  \cite{weyl1928, grossman1976} based on the Weyl-Heisenberg group. But this approach also includes more unusual examples. Indeed, because the symmetry group of the half-plane is the affine group and not the Weyl-Heisenberg one, IQ allows to develop a quantization of the half-plane based on the affine group that differs from the usual canonical prescription. When applied to singular cosmological models defined in the phase space which belongs to the half-plane, this difference between the canonical prescription and the affine quantization leads to major effects as regularization of the big-bang singularity. The semiclassical framework based on the affine CS allows to investigate the essential features of this quantum dynamics. It yields a semiclassical portrait in the phase space and exhibits a quantum corrections in the form of a repulsive potential which is responsible for replacing the classical big-bang singularity with a quantum bounce.\\ 
The problem of time is characteristic to gravitational systems. It refers to the lack of a fixed, external time. In order to describe the evolution of the gravitational field, one chooses an internal degree of freedom, the so-called internal time variable. The free choice of internal time is incompatible with ordinary quantum mechanics and leads to an unusual ambiguity of the respective quantum theory. The semiclassical framework based on affine CS proves very efficient in investigating this ambiguity and allows for better understanding of the nature of quantum evolution in gravity.\\
The paper is organized as follows. In Sec. \ref{sec:intq} we summarize the main features of the IQ framework and the special case of affine quantization. In Sec. \ref{sec3} we overview the canonical formulation of cosmological models and their quantization. In Sec. \ref{qMix} we develop a detailed analysis of the quantum Bianchi IX model. In Sec. \ref{QA} we analyze special features of the quantum Bianchi IX anisotropic Hamiltonian. In Sec. \ref{TI} we investigate the time problem. We conclude in Sec. \ref{C}.

\section{Integral quantization and coherent states}
\label{sec:intq}
Integral quantization \cite{hb2014, alijp2013,jpM2016,jpH2015,hb2017} is a generic name for approaches to quantization based on operator-valued measures. It includes the so-called Berezin-Klauder-Toeplitz quantization, and more generally coherent state quantization \cite{alijp2013, perelomov1986}. The integral quantization framework includes as well quantizations based on Lie groups. In the sequel we will refer to this case as \emph{covariant integral quantization}. The most famous example is the covariant integral quantization based on the Weyl-Heisenberg group (WH), like Weyl-Wigner \cite{weyl1928, grossman1976, daub1980a, daub1980b, daub1983} and (standard) coherent states quantizations \cite{perelomov1986}. It is well established that the WH group underlies the canonical commutation rule, a paradigm of quantum physics. Actually, there is a world of quantizations that follow this rule \cite{hb2014,hb2017}. 
This approach also includes a more unusual quantization of the half-plane based on the affine group \cite{hb2014, jpM2016}. The latter is essential in our approach of quantum cosmology \cite{QC2014, QC2015a, QC2015b, QC2016} developed below. Let us notice that  the affine group and related coherent states were also used for quantization of the half-plane in previous works by J.R. Klauder, although with a different approach (see \cite{klauder1970, klauder2011,fanuel2013} with references therein).

\subsection{General settings}
\label{sec:general}
Given a set $X$ and a vector space $\mathcal{C}(X)$ of complex-valued functions $f(x)$ on $X$, a quantization is a linear map  $\mathfrak{Q}: f\in  \mathcal{C}(X) \mapsto \mathfrak{Q}(f) \equiv A_f \in \mathcal{A}({\cal H})$ from $\mathcal{C}(X)$ to a vector space $\mathcal{A}({\cal H})$ of linear operators on some Hilbert space ${\cal H}$. Furthermore this map must fulfill the following conditions:\\
(i) To $f=1$ there corresponds $A_f = I_{\cal H}$, where $ I_{\cal H}$ is the identity in ${\cal H}$,\\
(ii) To a real function $f \in \mathcal{C}(X)$ there corresponds a(n) (essentially) self-adjoint operator $A_f$ in ${\cal H}$.\\
Physics puts into the game further requirements, depending on various mathematical structures allocated to $X$ and $\mathcal{C}(X)$, such as a measure, a topology, a manifold, a closure etc., together with an interpretation in terms of measurements. \\
Let us assume in the sequel that $X=G$ is a Lie group with left Haar measure ${\rm d}\mu(g)$, and let $g \mapsto U_g$ be a Unitary Irreducible Representation (UIR) of $G$ in a Hilbert space ${\cal H}$. Let $M$ be a bounded self-adjoint operator on ${\cal H}$ and let us define $g$-translations of $M$ as
\begin{equation}
M(g)= U_g M U_g^\dagger\,.
\end{equation}
Using Schur's Lemma, we prove \cite{hb2014} that there exists some real constant $c_M \in \mathbb{R}$ such that the following resolution of the identity holds (in the weak sense of bilinear forms)
\begin{equation}
\label{resolId}
 \int_G M(g) \frac{{\rm d} \mu(g)}{c_M} = I_{\mathcal{H}}  \,.
 \end{equation}
For instance, in the case of a square-integrable unitary irreducible representation $U: g \mapsto U_g$, let us pick a unit vector $| \psi \rangle$ for which $c_M = \int_G {\rm d}\mu(g) |\langle \psi | U_g \psi \rangle |^2 < \infty$, i.e $| \psi \rangle$ is an admissible unit vector for $U$. With $M = |\psi \rangle \langle \psi |$ the resolution of the identity (\ref{resolId}) provided by the family of states 
$| \psi_g \rangle = U_g | \psi \rangle$ reads
\begin{equation}
 \int_G |\psi_g \rangle \langle \psi_g | \frac{{\rm d} \mu(g)}{c_M} = I_{\mathcal{H}}  \,.
 \end{equation}
Vectors $| \psi_g \rangle$ are named (generalized) coherent states (or wavelet) for the group $G$. \\
 The equation (\ref{resolId}) provides an integral quantization of complex-valued functions on the group $G$ as follows
 \begin{equation}
 \label{quantiz}
 f \mapsto A_f = \int_G M(g) f(g) \frac{{\rm d} \mu(g)}{c_M} \,.
 \end{equation}
 Furthermore, this quantization is covariant in the sense that $U_g A_f U_g^\dagger = A_F$ where $F(g') = ({\cal U}_g f)(g') = f(g^{-1} g')$, i.e. ${\cal U}_g: f \mapsto F$ is the regular representation if $f \in L^2(G, {\rm d}\mu(g))$.\\
Let us notice that the operator-valued integral above (\ref{quantiz}) is understood in a weak sense, i.e. as the sesquilinear form
\begin{equation}
\mathcal{H} \ni \psi_1, \psi_2 \mapsto B_f(\psi_1,\psi_2) = \int_G \langle \psi_1| M_g | \psi_2 \rangle f(g) \frac{{\rm d} \mu(g)}{c_M} \,,
\end{equation}
where the form $B_f$ is assumed to be defined on a dense subspace of $\mathcal{H}$. If $f$ is a complex bounded function, $B_f$ is a bounded sesquilinear form, and from the Riesz lemma we deduce that there exists a unique bounded operator $A_f$ associated with $B_f$. If $f$ is real and semi-bounded, and if $M$ is a positive operator, Friedrich's extension of $B_f$ (\cite{reed}, Thm. X23) univocally defines a self-adjoint operator. However, if $f$ is real but not semi-bounded, there is no natural choice for a self-adjoint operator associated with $B_f$. In this case, we can consider directly the symmetric operator $A_f$ enabling us to obtain a possible self-adjoint extension (an example of this kind of mathematical study is presented in \cite{hb2012}). 

Integral quantization allows also to develop a natural semi-classical framework. If $\rho=M$  and $\tilde{\rho}$ are two positive unit trace operators, we obtain the exact classical-like expectation value formula
\begin{equation}
\label{semiclass1}
{\rm tr}(\tilde{\rho} A_f) = \int_G f(g) w(g) \frac{{\rm d} \mu(g)}{c_M} 
\end{equation}
where, up to the coefficient $c_M$, $w(g) = {\rm tr}(\tilde{\rho} M(g)) \ge 0$ is a classical probability distribution on the group. Furthermore we obtain a generalization of the Berezin or heat kernel transform on $G$:
\begin{equation}
\label{semiclass2}
f \mapsto \check{f}(g) = \int_G {\rm tr}(\tilde{\rho}_g \rho_{g'} ) f(g') \frac{{\rm d} \mu(g)}{c_M} 
\end{equation}
where $\tilde{\rho}_g \equiv M(g)$ when $M = \tilde{\rho}$ and $\rho_{g'} \equiv M(g')$ when $M = \rho$. The map $f \mapsto \check{f}$ is a generalization of the Segal-Bargmann transform \cite{stenzel1994}. Furthermore, the function or lower symbol $\check{f}$ may be viewed as a semi-classical representation of the operator $A_f$. In the case of coherent states $|\psi_g \rangle$ (i.e. $M=\rho = |\psi \rangle \langle \psi |$), Eq.(\ref{semiclass1}) reads
\begin{equation}
\label{semiclass3}
{\rm tr}(\tilde{\rho} A_f) = \int_G f(g) \, \langle \psi_g | \tilde{\rho} | \psi_g \rangle \frac{{\rm d}\mu(g)}{c_M}\,,
\end{equation}
where $w(g) = \langle \psi_g | \tilde{\rho} | \psi_g \rangle \ge 0$ acts as a classical probability distribution on the group (up to the coefficient $c_M$). Similarly assuming $\tilde{\rho} =  |\tilde{\psi} \rangle \langle \tilde{\psi} |$, the lower symbol $\check{f}(g)$ involved in (\ref{semiclass2}) reads
\begin{equation}
\label{semiclass4}
\check{f}(g) = \int_G |\langle \tilde{\psi}_g | \psi_{g'} \rangle |^2 f(g') \frac{{\rm d}\mu(g')}{c_M}
\end{equation}
This point will be developed at length in the case of the affine group.

\subsection{Half-plane and the affine group}
\subsubsection{Quantization of the half-plane}
\label{sec:affinegroup}
The half-plane is defined as $\Pi_+ = \{(q,p) \,|\, q>0, p \in \mathbb{R} \}$. Equipped with the law
\begin{equation}
(q,p)\cdot(q',p') = \left(q q', p+ \frac{p'}{q} \right) \,,
\end{equation}
$\Pi_+$ is viewed as the affine group Aff$_+(\mathbb{R})$ of the real line. The left invariant measure is  ${\rm d} \mu(q,p) = {\rm d}q {\rm d}p$. The group possesses two nonequivalent square integrable UIRs. Equivalent realizations of one of them, say, $U$, are carried on  Hilbert spaces $L^2(\mathbb{R}_+, {\rm d}x/x^\alpha)$. Nonetheless these multiple possibilities do not introduce noticeable differences. Therefore we choose in the sequel $\alpha=0$, and denote $\mathcal{H} = L^2(\mathbb{R}_+, {\rm d}x)$. The UIR of  Aff$_+(\mathbb{R})$ expressed in terms of the physical phase-space variables (q,p), acts on $\mathcal{H}$ as
\begin{equation}
U_{q,p} \psi(x) = \frac{1}{\sqrt{q}} e^{i p x} \psi(x/q) \,.
\end{equation}
Given a unit vector $\psi \in \mathcal{H}$, we define the Affine Coherent States (ACS) as follows
\begin{equation}
|\psi_{q,p} \rangle = U_{q,p} | \psi \rangle\,,
\end{equation}
where $\psi$ is called the fiducial vector. Using the framework of covariant integral quantization presented above, we first notice that the following resolution of the identity holds
\begin{equation}
\int_{\Pi_+}  |\psi_{q,p} \rangle \langle \psi_{q,p} |\,  \frac{{\rm d}q {\rm d}p}{2 \pi c} = I_{\mathcal{H}} \,,
\end{equation}
provided that $c = \int_0^\infty |\psi(x)|^2 {\rm d}x /x < \infty$. Therefore the covariant integral quantization follows:
\begin{equation}
\label{affineq}
f \mapsto A_f = \int_{\Pi_+} f(q,p) \, |\psi_{q,p} \rangle \langle \psi_{q,p} | \,\, \frac{{\rm d}q {\rm d}p}{2 \pi c}
\end{equation}
Note that the idea of using in quantum gravity an ÔÔaffineÕÕ quantization instead of the Weyl-Heisenberg one was already present in Klauder's work \cite{klauder1970} devoted to the question of singularities in quantum gravity (see \cite{klauder2011} for recent references). The procedure followed by Klauder  rests on the representation of the affine Lie algebra. In this sense, it remains closer to the canonical one and it is not of the integral type. 

 In the sequel let us assume without loss that the fiducial function $\psi$ is a \underline{real} function of rapid decrease on $\mathbb{R}_+$. This ensures the convergence of the different integrals $c_\alpha$ defined as
 \begin{equation}
 c_\alpha = \int_0^\infty \frac{{\rm d}x}{x^{2+\alpha}} \psi(x)^2\,.
 \end{equation}
Note that the coefficient $c$ involved in (\ref{affineq}) reads $c \equiv c_{-1}$, and the normalization of $\psi$ corresponds to $c_{-2} = 1$. 

The first interesting issue of the map (\ref{affineq}) is that the quantization yields canonical commutation
rule, up to a scaling factor, for $A_q$ and $A_p$:
\begin{equation}
A_p = P = - i \frac{{\rm d}}{{\rm d}x}, \quad A_q = (c_0/c_{-1}) Q, \quad Q \psi(x)
=  x \psi(x), \quad [A_q, A_p] = i \frac{c_0}{c_{-1}} I_{\mathcal{H}}
\end{equation}
By a unitary rescaling of the fiducial vector $\psi(x) \mapsto \lambda^{-1/2} \psi(x/\lambda)$ with $\lambda = c_0/c_{-1}$ we can impose $c_0 =c_{-1}$ and then recover the usual canonical rule. To simplify expressions we assume this condition to be fulfilled in the sequel.\\
However, while $A_q=Q$ is (essentially) self-adjoint, we know from \cite{reed} that $A_p=P$ is symmetric but has no self-adjoint extension. The quantization of any power of $q$ is canonical , up to a scaling factor:
\begin{equation}
A_{q^\beta} = \frac{c_{\beta-1}}{c_{-1}} Q^\beta\,.
\end{equation}
Note that our assumption on the rapid decrease of $\psi$ ensures the finiteness of the coefficients $c_{\beta-1}$, whatever $\beta$.\\
The quantization of the product $qp$ yields
\begin{equation}
A_{qp} = \frac{1}{2} (Q P + P Q) \equiv D\,,
\end{equation}
where $D$ is the dilation generator. As one of the two generators (with $Q$) of the UIR $U$ of the affine group, it is essentially self adjoint.\\
The last and the main result  is a regularization of the quantum ``kinetic energy'':
\begin{equation}
\label{kinetic}
A_{p^2} = P^2+ \frak{K}_{\psi} \, Q^{-2} \quad \textrm{with} \quad \frak{K}_{\psi} = \int_0^\infty \frac{{\rm d}u}{c_{-1}} \, u \, (\psi'(u))^2.
\end{equation}
Therefore this quantization procedure yields a non-canonical additional term. This term is a centrifugal potential whose strength depends on the fiducial vector only. In other words, this affine quantization forbids a quantum free particle moving on the positive line to reach the origin. Now, it is known \cite{reed, gesztesy1985} that the operator $P^2 = - {\rm d}^2 / {\rm d}x^2$ alone in $L^2(\mathbb{R}^+, {\rm d}x)$ is not essentially self-adjoint whereas the regularized operator (\ref{kinetic}) is for $\frak{K}_{\psi} \ge 3/4$. It follows that for $\frak{K}_{\psi} \ge 3/4$ the quantum dynamics is unitary during the entire evolution, in particular in the passage from the motion towards $Q = 0$ to the motion away from $Q = 0$. 

\subsubsection{Semiclassical framework}
\label{sec:sIntemiclass}
The semiclassical framework sketched in section (\ref{sec:general}) applies naturally for the half-plane viewed as the affine group. The quantum states and their dynamics have phase space representations through wavelet symbols. For a state $|\phi \rangle$ one has the associated probability distribution $\rho_\phi(q,p)$ on phase space given by the substitution $\tilde{\rho} = | \phi \rangle \langle \phi|$ in (\ref{semiclass3})
\begin{equation}
\rho_\phi(q,p) = \frac{1}{2\pi c_{-1}} | \langle \psi_{q,p} | \phi \rangle |^2
\end{equation}
To apply the map (\ref{semiclass4}) yielding lower symbols from classical $f$ we introduce two different real fiducial functions $\psi$ and $\tilde{\psi}$. The vector $\psi$ is devoted to quantization and submitted to the constraints $c_{-2} = 1$, $c_0 = c_{-1}$, while $\tilde{\psi}$ is only constrained by the normalization $\tilde{c}_{-2}=1$. The map (\ref{semiclass4}) reads in the present case:
\begin{align}
\nonumber \check{f}(q,p) = &\frac{1}{\sqrt{2\pi} c_{-1}} \int_0^\infty \frac{{\rm d}q'}{q q'} \int_0^\infty {\rm d}x \int_0^\infty {\rm d}x' e^{i p (x'-x)} \\
\times & F_{\frak{p}}(q',x-x') \tilde{\psi} \left(\frac{x}{q} \right) \tilde{\psi} \left(\frac{x'}{q} \right) \psi \left(\frac{x}{q'} \right) \psi \left(\frac{x'}{q'} \right)\,,
\end{align}
where $F_{\frak{p}}$ stands for the partial inverse Fourier transform
\begin{equation}
F_{\frak{p}}(q,x) = \frac{1}{\sqrt{2\pi}} \int_{-\infty}^{+\infty} e^{i p x} \, f(q,p) {\rm d}p \,.
\end{equation}
For instance, any power of $q$ is transformed into the same power up to a constant factor
\begin{equation}
f(q,p) = q^\beta \mapsto \check{f}(q,p) = \frac{\tilde{c}_{-\beta-2} \, c_{\beta-1}}{c_{-1}} q^\beta \, ,
\end{equation}
where $\tilde{c}$ coefficients stand for $\tilde{\psi}$. \\
We notice that $\check{q} = c_0\, \tilde{c}_{-3} \, (c_{-1})^{-1} q = \tilde{c}_{-3} \, q$. Therefore we must impose $\tilde{c}_{-3} = 1$ if we want to obtain for physical consistency $\check{q} = q$. This constraint is obtained by a simple rescaling of the fiducial vector $\tilde{\psi}$. We assume this condition to be fulfilled in the sequel.\\
Other important symbols are
\begin{align}
f(q,p) = p & \mapsto \check{f}(q,p) = p\\ \label{LSofHAM}
f(q,p) =p^2 &\mapsto \check{f}(q,p) = p^2+ \frak{K}_s(\tilde{\psi},\psi) \, q^{-2}\\
f(q,p) = q p & \mapsto \check{f}(q,p) = q p
\end{align}
where
\begin{equation}\label{SofKs}
\frak{K}_s(\tilde{\psi},\psi) = \int_0^\infty (\tilde{\psi}'(u))^2 \, {\rm d}u + \tilde{c}_0\, \frak{K}_\psi \,.
\end{equation}

\subsection{Weyl-Heisenberg integral quantization}

We start with the homogeneity of the plane, where the choice of the origin is arbitrary. We then impose our quantization to be covariant with respect to this basic symmetry (\cite{JPGip}). This leads to the integral quantization which transforms a function $f(q,p) \equiv f(\vec{r})$ into an operator $A_f$ in some Hilbert space $\mathcal{H}$ through the linear map
\begin{equation} \label{fAf}
f(\vec{r}) \mapsto A_f = \int_{\mathbb{R}^2} f(\vec{r})\, \mathfrak{Q}(\vec{r})\, \frac{\mathrm{d}^2\vec{r}}{2\pi c_{\mathfrak{Q}}}\, ,\quad \mathrm{d}^2 \vec{r}=\mathrm{d} q\,\mathrm{d} p\, . 
\end{equation}
where $c_{\mathfrak{Q}}$ is some positive constant and  $\mathfrak{Q}(\vec{r})/c_{\mathfrak{Q}}$ is a family of operators in $\mathcal{H}$  which solve the identity:
\begin{equation}
\label{1A1}
\int_{\mathbb{R}^2}  \mathfrak{Q}(\vec{r})\, \frac{\mathrm{d}^2\vec{r}}{2\pi c_{\mathfrak{Q}}} = I\,.
\end{equation}
Translational covariance should hold in the sense that the quantization of the translation of $f$ is unitarily equivalent to the quantization of $f$  as follows:
\begin{equation} \label{covtrans1}
U(\vec{r}_0)\,A_f \,U(\vec{r}_0)^{\dag}= A_{\mathcal{T}(\vec{r}_0)f}\, , \quad \left(\mathcal{T}(\vec{r}_0)f\right)(\vec{r}):= f\left(\vec{r}-\vec{r}_0\right) \, . 
\end{equation} 
 So $\vec{r}\mapsto U(\vec{r})$ has to be a unitary {projective  representation} of the abelian group $\mathbb{R}^2$. This leads naturally to the unique (up to equivalence) Weyl-Heisenberg representation :
\begin{eqnarray}
\label{WHU}
U(\vec{\mathbf{0}}) &= I\, , \quad U^{\dag}(\vec{r}) = U(-\vec{r})\, , \\
 U(\vec{r})\,U(\vec{r}p)& = e^{\mathrm{i} \xi(\vec{r},\vec{r}p)}U(\vec{r}+\vec{r}p)\, , 
\end{eqnarray}
where the real valued $\xi$ encodes the non commutativity of the representation which is the central feature of the quantum $A_f$. It has   to fulfill  cocycle conditions which correspond with group structure of $\mathbb{R}^2$. Therefore $\xi(\vec{r},\vec{r}') $ is  bilinear in $(\vec{r},\vec{r}')$. From $\xi(\vec{r},-\vec{r}) = - \xi(\vec{r},\vec{r})= 0$ there follows that the only possibility is that $U(\vec{r}) =  e^{\mathrm{i} (p\hat q-q\hat p)}$ is the unitary displacement operator and $\xi(\vec{r},\vec{r}') $ is the symplectic form:
\begin{equation}
\xi(\vec{r},\vec{r}p) = k\,(qp^{\prime}-q^{\prime}p)\equiv k\,\vec{r}\wedge\vec{r}'\, . 
\end{equation}
Here  $k$ is a parameter that quantum physics fixes to $1/\hbar$, and  for convenience it is  put equal to $1$ in these considerations.
 Then, from \eqref{covtrans1} and the translational invariance of $\mathrm{d}^2\vec{r}=\mathrm{d} q\,\mathrm{d} p$, the operator valued function  $\mathfrak{Q}(\vec{r})$ has to obey 
\begin{equation}
\label{covQ}
U(\vec{r}_0)\,\mathfrak{Q}(\vec{r}) \,U(\vec{r}_0)^{\dag}= \mathfrak{Q}\left(\vec{r} + \vec{r}_0\right) \,. 
\end{equation}
The solution to \eqref{covQ} is  easily found by picking an operator $\mathfrak{Q}_0\equiv \mathfrak{Q}(\vec{\boldmath{0}})$ and reads 
\begin{equation}
\label{solQ}
\mathfrak{Q}\left(\vec{r}\right) := U(\vec{r})\,\mathfrak{Q}_0 \,U(\vec{r})^{\dag} \, .
\end{equation}
The choice of  $\mathfrak{Q}_0$ is admissible provided that  $0<c_{\mathfrak{Q}_0}< \infty$, and if $\mathfrak{Q}_0$ is trace class, i.e. for finite $\mathrm{Tr}(\mathfrak{Q}_0)$.

Let us now introduce  the ``WH-transform'' of the operator $\mathfrak{Q}_0$ and its inverse as follows
\begin{equation}
\label{WHT}
\Pi(\vec{r}) = \mathrm{Tr}\left(U(-\vec{r})\mathfrak{Q}_0 \right)\  \Leftrightarrow\ \mathfrak{Q}_0 = \int_{\mathbb{R}^2} U(\vec{r}) \, \Pi(\vec{r})\,\frac{\mathrm{d}^2\vec{r}}{2\pi}\,. 
\end{equation}
The inverse WH-transform exists due to remarkable properties of the displacement operator $U(\vec{r})$:
\begin{equation}
\label{IWHT}
 \int_{\mathbb{R}^2} U(\vec{r}) \,\frac{\mathrm{d}^2\vec{r}}{2\pi}= 2{\sf P}\ \mbox{and}\ \mathrm{Tr}\left(U(\vec{r})\right)= 2\pi \delta(\vec{r}) \ \rightarrow \
\mathfrak{Q}_0 = \int_{\mathbb{R}^2} U(\vec{r}) \, \Pi(\vec{r})\,\frac{\mathrm{d}^2\vec{r}}{2\pi}\, ,
\end{equation}
where ${\sf P}= {\sf P}^{-1}$ is the parity operator defined as ${\sf P}U(\vec{r}){\sf P}= U(-\vec{r})$. 
 The value of constant $c_{\mathfrak{Q}_0}$ can be derived as 
\begin{equation}
\label{cQ0}
c_{\mathfrak{Q}_0} = \mathrm{Tr}\left(\mathfrak{Q}_0 \right) = \Pi(\vec{\mathbf{0}})\,. 
\end{equation} 
We have at our disposal also an alternative integral quantization formula through the so-called symplectic Fourier transform:
 \begin{equation}
\label{symFourqp}
 \mathfrak{F_s}[f](\vec{r})= \int_{\mathbb{R}^2}e^{-\mathrm{i} \vec{r}\wedge\vec{r}p}\, f(\vec{r}')\,\frac{\mathrm{d}^2\vec{r}'}{2\pi} \, . 
\end{equation}
 It is involutive, $\mathfrak{F_s}\left[\mathfrak{F_s}[f]\right]=  f$ is defined as $\overline{\mathfrak{F_s}}[f](\vec{r})= \mathfrak{F_s}[f](-\vec{r})$. Hence the equivalent form of  the WH integral quantization:
 \begin{equation} 
 \label{SIQ}
A_f= \int_{\mathbb{R}^2}  U(\vec{r})\,  \overline{\mathfrak{F_s}}[f](\vec{r})\, \frac{\Pi(\vec{r})}{\Pi(\vec{\mathbf{0}})} \,\frac{\mathrm{d}^2 \vec{r}}{2\pi}\,.  
\end{equation}
There are several features  independent of the choice of the quantization operator $\mathfrak{Q}_0$.
First, the canonical commutation rule is preserved
\begin{equation}
\label{AqAp}
A_q = \hat q + c_0\, , \quad A_p= \hat p+d_0\,, \quad c_0,d_0\in \mathbb{R}\, ,  \rightarrow \left[A_q,A_p\right]= \mathrm{i} I\, .
\end{equation}
For the kinetic energy we have the following formula
\begin{equation}
 A_{p^2}= \hat p^2 + e_1\,\hat p + e_0\, , \quad e_0, e_1 \in \mathbb{R}\, . 
\end{equation}
The constants $c_0$, $d_0$, $e_0$, $e_1$ appearing in the above may vanish with a suitable choice of $\mathfrak{Q}_0$.
The quantization of the dilatation  operator yields:
\begin{equation}
A_{qp} = A_q\,A_p + \mathrm{i} f_0\, , \quad  f_0\in \mathbb{R}\, . 
\end{equation}
This operator can be brought to the self-adjoint dilation operator $(\hat q\hat p + \hat p\hat q)/2$ again with a suitable choice of $\mathfrak{Q}_0$.

A potential energy becomes the multiplication operator in the position representation
\begin{equation}
\label{Iqvq}
A_{V(q)} = \mathfrak{V}(\hat q)\, , \quad \mathfrak{V}(\hat q)= \frac{1}{\sqrt{2\pi}}\,V\ast \overline{\mathcal{F}}[\Pi(0,\cdot)](\hat q)\,
\end{equation}
where $ \overline{\mathcal{F}}$ is the inverse $1$-$d$ Fourier transform, and $f\ast g(x)= \int_{\mathbb{R}}\mathrm{d} t\, f(x-t)\,g(t)$. Such a convolution formula can be of crucial importance when it is needed to smooth classical singularities or modify in a suitable way the strengths of some potentials as will be shown in the sequel.  

 Finally, if $F(\vec{r})\equiv h(p)$ is a function of $p$ only, then $A_h$ depends on $\hat p$ only
\begin{equation}
\label{Iqvp}
A_h= \frac{1}{\sqrt{2\pi}}\,h\ast \overline{\mathcal{F}}[\Pi(\cdot,0)](\hat p)\, .
\end{equation}

\section{Quantization of cosmological models}\label{sec3}
The phase space formalism of general relativity was introduced by Arnowitt, Deser and Misner in \cite{adm}. The main features of their formalism are (i) an ambiguous split of the spacetime into a spatial leaf and a time manifold, (ii) the non-linearity and (iii) the appearance of four constraints. The constraints play two roles: on the one hand, they confine the physically admissible states to a submanifold in the phase space and on the other hand, they generate canonical transformations which are interpreted as spacetime diffeomorphisms. The so-called vector constraints generate spatial diffeomorphisms and thus, they represent pure gauge transformations. On the other hand, the so-called scalar constraint generates a diffeomorphism directed in the time-like normal to the space-like leaf and includes dynamics. Therefore, the Hamiltonian in canonical relativity is a constraint. This is an expression of the lack of a fixed, external time in the theory and is a starting point for the discussion of the time problem later on in the present article.

\subsection{Canonical formulation of spatially homogenous models}
The so-called class A Bianchi type models are a family of relativistic cosmological models that admit a three-parameter group of symmetry in the spatial leaf generated by three independent Killing vectors satisfying the following algebra,
\begin{equation}
\{\xi_j,\xi_k\}=C_{jk}^i\xi_i,~~C^{jk}_i=\epsilon_{jkl}h^{li},
\end{equation}
where $h^{li}$ is a symmetric matrix. Suppose $h^{li}$ is diagonal and $\omega_i$: $\omega_i(\xi_j)=\delta_{ij}$ are dual 1-forms. Then the line element of the diagonal class of respective metrics reads \cite{Misner69}:
\begin{equation}\label{misner_metrics}
\ud s^2=-N^2\ud t^2+e^{2(\beta_0+\beta_++\sqrt{3}\beta_-)}\omega_1^2+e^{2(\beta_0+\beta_+-\sqrt{3}\beta_-)}\omega_2^2+e^{2(\beta_0-2\beta_+)}\omega_3^2,
\end{equation}
where the lapse function $N$, the isotropic variable $\beta_0$ and the anisotropic variables $\beta_{\pm}$ depend only on time $t$. 

The scalar constraint for the symmetry-reduced class of metrics (\ref{misner_metrics}) reads \cite{uggla}
\begin{equation}\label{scalar}
\mathrm{C}=\frac{Ne^{-3\beta_0}}{24}\left(-p_0^2+p_+^2+p_-^2+24e^{4\beta_0}V_M(\beta_{\pm})+24e^{3(1-w)\beta_0}p_T\right),
\end{equation}
where $V_M(\beta_{\pm})$ depends on the particular model denoted by the number $M$. The momenta $p_0$ and $p_{\pm}$ are canonically conjugate to the metric variables $\beta_0$ and $\beta_{\pm}$, respectively. The momentum $p_T>0$ is a momentum conjugate to $T$ and associated with a perfect fluid satisfying $$w=\frac{pressure}{energy~ density}=const.$$ Note that the vector constraints of the full canonical formalism identically vanish in the Bianchi models case as the dynamics is orthogonal to the spatial leaf, which is visible from the form of the metric (\ref{misner_metrics}). 

The lapse function $N$ is arbitrary and its choice determines the time parameter $t$. For the choice $N=e^{3w\beta_0}$, the scalar constraint (\ref{scalar}) becomes linear with respect to $p_T$, i.e. the constraint (\ref{scalar}) acquires the following form,
\begin{equation}
\mathrm{C}=p_T+\mathrm{H}(\beta_0,p_0,\beta_{\pm},p_{\pm},T).
\end{equation}
It follows that the variable $T$ may be identified with the internal time variable by removing the pair $(T,p_T)$ from the initial phase space and reducing it to the so called reduced phase space given by $(\beta_0,p_0,\beta_{\pm},p_{\pm})$. We keep the commutation rules between those variables and introduce the so called true Hamiltonian,
\begin{equation}
\mathrm{H}=\mathrm{H}(\beta_0,p_0,\beta_{\pm},p_{\pm},T),
\end{equation}
which generates the physical motion in the time variable $T$. The momentum $p_T$ is a redundant quantity absent in the reduced formalism. This approach is sometimes called ``deparametrization" or ``reduced phase space approach" and it is discussed e.g. in \cite{Ish}.

Note that at the big-bang/big-crunch singularity the volume of the universe $V=e^{3\beta_0}$ vanishes and thus, $\beta_0\rightarrow -\infty$. From the phase space formalism perspective, the singularity is localized at infinity and thus, hidden. In order to make its existence more apparent, one redefines the canonical pair of isotropic variables $(\beta_0,p_0)$,
\begin{equation}
p=e^{-\frac{3}{2}(1-w)\beta_0}p_0,~~~q=\frac{2}{3(1-w)}e^{\frac{3}{2}(1-w)\beta_0},
\end{equation}
where $q>0$ and the singularity occurs at a finite distance for $q=0$. For the vacuum case, $p_T=0$, an analogous redefinition of variables also holds.
\subsection{Quantum and semiclassical Friedmann-Lemaitre model}
 Let us briefly discuss an example of the flat Friedmann (FRW) model \cite{QC2014} for which the anisotropic variables vanish, $\beta_{\pm}=0=p_{\pm}$, and so does the potential in the scalar constraint (\ref{scalar}), i.e. $V_{I}(\beta_{\pm})=0$. For a convenient choice of $N$, the scalar constraint reads,
\begin{equation}
\mathrm{C}=-\frac{1}{24}p^2+p_T.
\end{equation}
Solving the constraint with respect to $p_T$ removes the fluid variables from the phase space and establishes the true Hamiltonian of the reduced phase space formalism,
\begin{equation}\label{FRWH}
\mathrm{H}=\frac{1}{24}p^2.
\end{equation}
The above Hamiltonian describes a free motion of a particle in the half-line, $q>0$. As showed by Eq. (\ref{kinetic}), the affine coherent state quantization of the Hamiltonian  gives,
\begin{equation}\label{freep}
\mathrm{H}\mapsto A_{\mathrm{H}}=\frac{1}{24}\left(P^2+\frac{\frak{K}_{\psi}}{Q^2}\right),
\end{equation}
where the value of $\frak{K}_{\psi}$ depends on the fiducial vector $\psi$. We note that the quantum Hamiltonian (\ref{freep}), which is defined on the half-line, is (essentially) self-adjoint for $\frak{K}_{\psi}\ge\frac{3}{4}$. The role of the new term $\propto \frac{1}{Q^2}$ is to produce a repulsive force that prevents the particle from reaching the origin point $Q=0$. Geometrically, the singular state of vanishing volume is shielded by a quantum repulsive force issued by the affine quantization. 

\begin{figure}[!t]
\includegraphics[width=0.45\textwidth]{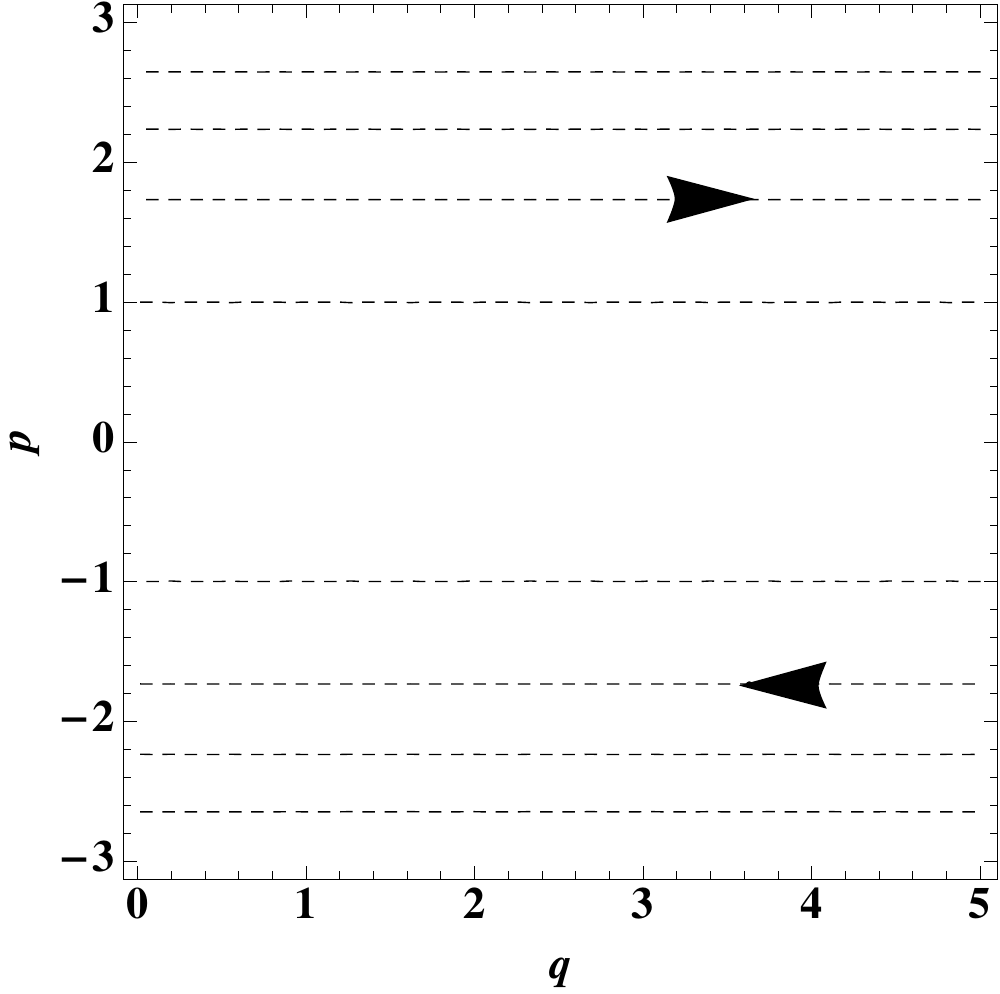}
\hspace{1cm}
\includegraphics[width=0.45\textwidth]{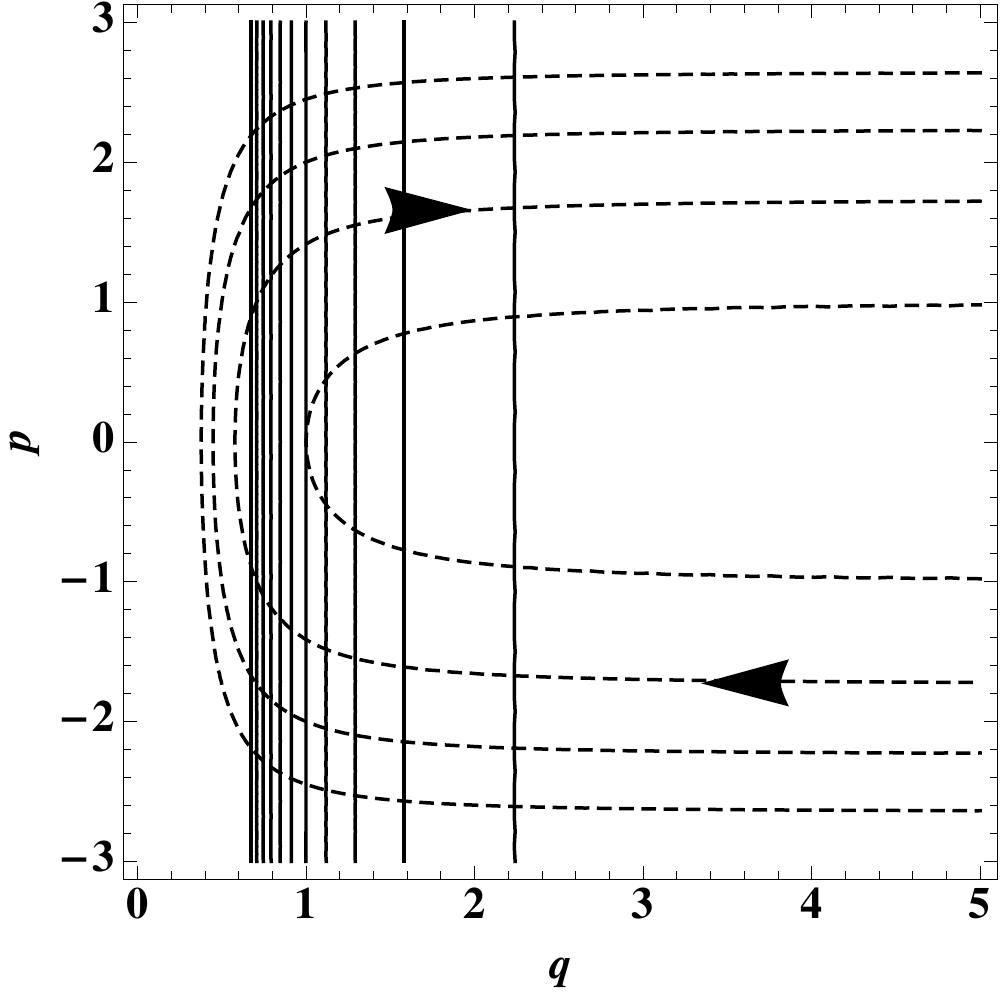}
\caption{On the left: The classical Friedman phase space trajectories terminate in the $q=0$ singularity. On the right: The semiclassical Friedman trajectories are generated by a semiclassical Hamiltonian that includes a semiclassical correction, the repulsive potential with $\frak{K}_s(\tilde{\psi},\psi)=1$. The vertical lines are the equipotential lines of the repulsive quantum-induced potential which produces the bounces. (Source: \cite{M3})}
\label{fig0}
\end{figure}

The quantum dynamics can be approximated by confining the quantum motion to a family of the affine coherent states  \cite{klauder2012} that we construct with another fiducial vector, say $\tilde{\psi}$. It can be shown that the resultant motion in terms of the canonical variables is generated by the lower symbol of the Hamiltonian (\ref{FRWH}). Hence, according to Eq. (\ref{LSofHAM}),
\begin{equation}
\frac{\ud q}{\ud T}=\{q,\check{\mathrm{H}}\},~~\frac{\ud p}{\ud T}=\{p,\check{\mathrm{H}}\},~~\check{\mathrm{H}}=p^2+\frac{\frak{K}_s(\tilde{\psi},\psi)}{q^2},
\end{equation}
where $\frak{K}_s(\tilde{\psi},\psi)$ is given in Eq. (\ref{SofKs}). The semiclassical dynamics is illustrated in Fig. \ref{fig0} with several solutions. 

The discussed flat Friedman model can be extended to the anisotropic Bianchi I model with non-vanishing $(\beta_{\pm},p_{\pm})$. The respective affine coherent state quantization and semiclassical formalism can be found in \cite{BI}. The application of the presented methods to the vacuum Bianchi IX model is discussed in the next section.

\section{Resolution to the Mixmaster singularity}\label{qMix}
\begin{figure}[!ht]
\sidecaption
\includegraphics[width=0.45\textwidth]{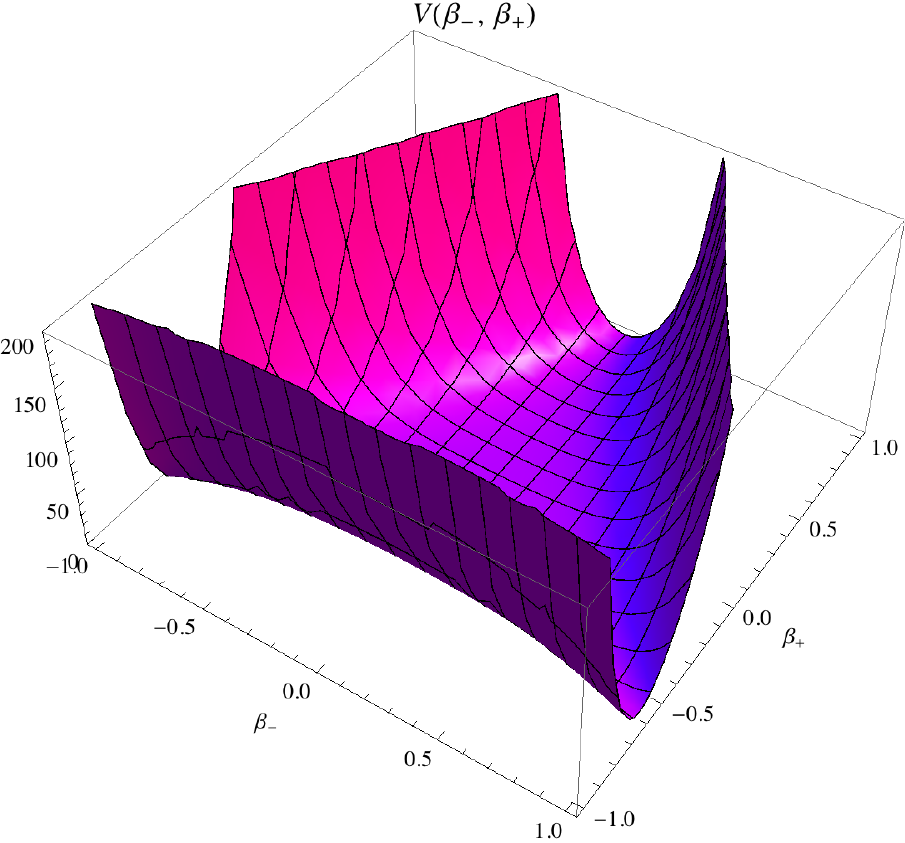}
\caption{Plot of the Bianchi type IX anisotropy potential $V_{IX}(\beta_+, \beta_-)$ near its minimum with the three ${\sf C}_{3v}$ symmetry axes $\beta_-=0$, $\beta_+=\beta_-/\sqrt{3}$, $\beta_+=-\beta_-/\sqrt{3}$. (Source: \cite{QC2015a})}
\label{figBIXpot}
\end{figure}

The Mixmaster universe is a model of the spatially homogeneous and anisotropic spacetime that admits the Bianchi type IX symmetry. It exemplifies generic features of the oscillatory singularity driven by the gravitational self-energy. In the context of quantum gravity, the Mixmaster universe is ideal for testing whether quantization can resolve the problem of classical singularities.\\
We present in this section a quantum model that illustrates the interest of ACS integral quantization \cite{QC2015a, QC2015b, QC2016}. Our main purpose is to prove that the classical singularity is cured thanks to a repulsive potential generated by our affine quantization: the singularity is replaced by a bounce. \\
The full classical Hamiltonian of the vacuum Bianchi IX model involving the isotropic variable $q=a^{3/2}$ ($a$ is the scale factor) and anisotropy Misner variables $(\beta_+, \beta_-)$ reads (up to some physical constants chosen as units)
\begin{equation}
 \left.\begin{array}{c}
h = N\, \mathrm{C}; \quad \mathrm{C} =  \mathrm{C}^{{\rm (iso)}} - \mathrm{C}^{{\rm (anis)}}_q \\
 \mathrm{C}^{{\rm (iso)}} = \frac{9}{4} p^2+ 36 q^{2/3} \\
\mathrm{C}^{{\rm (anis)}}_q = \frac{1}{q^2} \left( p_+^2+p_-^2 \right) + 12 q^{2/3} V_{IX}(\beta_+, \beta_-)
\end{array}\right.
\end{equation}
where $N$ is the lapse, $\mathrm{C}^{{\rm (iso)}}$ is the isotropic part of the constraint $\mathrm{C}$ with  $(q,p) \in \Pi_+$ (canonical isotropic variables), and $\mathrm{C}^{{\rm (anis)}}_q$ is the anisotropic part of the constraint with $(\beta_\pm, p_\pm) \in \mathbb{R} \times \mathbb{R}$ (anisotropic canonical variables). The potential $V_{IX}(\beta_+,\beta_-)$ is the Bianchi IX anisotropy potential shown in Fig. \ref{figBIXpot}. It reads
\begin{equation}\label{eqB9pot}
V_{IX}(\beta_+, \beta_-) = \frac{e^{4\beta_+}}{3} \left[\left(2\cosh(2\sqrt{3}\beta_-)-e^{- 6\beta_+}
\right)^2-4\right] +  1 \,.
\end{equation}

\subsection{Quantum Bianchi IX model}
The quantum model is based on four main elements: 
\begin{itemize}
\item[(a)] A compound quantization procedure that fully complies with the symmetries of the phase space: an ACS quantization for isotropic variable which is consistent  with the dilation-translation symmetry of the half-plane (affine group), and a Weyl-Wigner quantization for anisotropic variables consistent with the translation symmetry of the plane (Weyl-Heisenberg group).
\item[(b)]  Inspired by Klauder's work about \emph{Enhanced quantization} \cite{klauder2012}, we develop a compound semi-classical Lagrangian approach: semiclassical for isotropic variable and purely quantum for anisotropy variables.
\item[(c)] Following standard approaches in molecular physics, we study successively adiabatic (Born-Oppenheimer-like) and nonadiabatic (vibronic-like) approximations. 
\item[(d)] We expand the anisotropy potential about its minimum in order to deal with its harmonic approximation suitable for both analytical and numerical treatments.
\end{itemize}
As noticed in Sec. \ref{sec:intq} the affine group and related coherent states were also used for quantization in previous Klauder's works, although with a different approach (see \cite{klauder1970, klauder2011,fanuel2013} with references therein).\\
We use the ACS quantization framework presented above for the isotropic pair $(q,p)$, and  a canonical quantization for the anisotropic pairs $(\beta_\pm, p_\pm)$. We obtain the quantized version $\hat{\mathsf{H}} \equiv A_h = N \, A_{\mathrm{C}}$ of the classical Hamiltonian $h$ acting on the Hilbert space $\mathcal{H} = \mathcal{H}^{{\rm (iso)}} \otimes  \mathcal{H}^{{\rm (anis)}}$ , where $\mathcal{H}^{{\rm (iso)}} =L^2(\mathbb{R}_+, {\rm d}x)$ and $\mathcal{H}^{{\rm (anis)}} =L^2(\mathbb{R}^2, {\rm d}\beta_+ {\rm d} \beta_-)$:
\begin{equation}
\label{qb9}
\left.\begin{array}{c}
\mathrm{C} \mapsto A_{\mathrm{C}} \equiv \hat{\mathrm{C}} =  \hat{\mathrm{C}}^{{\rm (iso)}} - \hat{\mathrm{C}}^{{\rm (anis)}}(Q)\\
 \hat{\mathrm{C}}^{{\rm (iso)}} \equiv A_{\mathrm{C}^{{\rm (iso)}}} = \frac{9}{4} \left(P^2+\frac{\frak{K}_1}{Q^2} \right)+ 36 \frak{K}_3 Q^{2/3}\\
\hat{\mathrm{C}}^{(\mathrm{anis})}(q) \equiv A_{\mathrm{C}^{({\sf anis})}_q}=\frak{K}_2\frac{\hat{p}_+^2+\hat{p}_-^2}{q^2}+
36 \frak{K}_3 q^{2/3} V_{IX}(\hat{\beta}_+,\hat{\beta}_-)\,,
\end{array}\right.
\end{equation}
where $\hat{p}_\pm = -i \partial_{\beta_\pm}$, and the coefficients $\frak{K}_1$, $\frak{K}_2$, $\frak{K}_3$ result from our ACS quantization, being only dependent on the ACS fiducial vector. \\
We recover the main interest of our ACS quantization pointed out in the section \ref{sec:affinegroup}, namely the creation of a repulsive potential $\frak{K}_1 Q^{-2}$ which will be responsible,  in the Bianchi IX framework,  of the resolution of the singularity.\\
Furthermore, the study of the Hamiltonian $\hat{\mathrm{C}}^{(\mathrm{anis})}(q)$ shows  that despite three open canyons, the potential $V_{IX}(\beta_+,\beta_-)$ originates a purely discrete spectrum \cite{QC2017}. Therefore $\hat{\mathrm{C}}^{(\mathrm{anis})}(q)$ possesses the discrete spectral resolution
\begin{equation}
\label{spectralres}
\hat{\mathrm{C}}^{(\mathrm{anis})}(q)  = \sum_n E_n^{{\rm (anis)}} (q) | e_n^{{\rm (anis)}} (q) \rangle \langle e_n^{{\rm (anis)}} (q) | \,.
\end{equation}
We prove in \cite{QC2017} that the eigenenergies $E_n(q)$ verify
$
\lim_{q \to 0} q^2 \, E_n(q) = 0\,.
$
Finally we introduce a unitary transformation $U(q,q')$ and a new self-adjoint operator $\hat{\mathsf{A}}(q)$ acting on the Hilbert space $\mathcal{H}^{{\rm (anis)}}$. They will be useful for the section below:
\begin{equation}
\label{newunitary}
U(q,q') = \sum_n | e_n^{{\rm (anis)}} (q) \rangle \langle e_n^{{\rm (anis)}} (q') | \,.
\end{equation}
We notice that $U(q,q')^\dagger = U(q',q)$ and $U(q,q) = I_{\mathcal{H}^{{\rm (int)}}}$.
We also define  the self-adjoint operator $\hat{\mathsf{A}}(q)$   as
\begin{equation}
\label{Afield}
\hat{\mathsf{A}}(q) = i \sum_n \left( \frac{\partial}{\partial q}  | e_n^{{\rm (anis)}} (q) \rangle \right)  \langle e_n^{{\rm (anis)}} (q)| = i \left( \frac{\partial}{\partial q} U(q,q') \right) U(q',q) \,.
\end{equation}

\subsection{Semiclassical formalisms}
We recall in this section our procedure detailed in \cite{QC2015b, QC2016}. It is inspired by Klauder's approach \cite{klauder2012} and is based on a consistent framework allowing us to approximate the quantum Hamiltonian and its associated dynamics (in the constraint surface) by making use of the semiclassical Lagrangian approach, which is made possible with the use of our ACS formalism. The quantum constraint (\ref{qb9}) has the general form
\begin{equation}
\left.\begin{array}{c}
\hat{\mathrm{C}} =\hat{\mathrm{C}}^{(\mathrm{iso})} - \hat{\mathrm{C}}^{(\mathrm{anis})}(Q)\\
\hat{\mathrm{C}}^{(\mathrm{iso})} =  \frac{9}{4}P^2+W(Q), \quad W(q) = \frac{9\frak{K}_1}{4 q^2} + 36 \frak{K}_3 q^{2/3}
\end{array}\right.
\end{equation}
and the $q$-dependent Hamiltonian $\hat{\mathrm{C}}^{(\mathrm{anis})}(q)$ is formally the one of (\ref{spectralres}) that acts on the
Hilbert space of anisotropy states. The Schr\"{o}dinger equation (here $\hbar=1$)
\begin{equation}
i \frac{\partial}{\partial t} | \Phi(t) \rangle = N \hat{\mathrm{C}} |\Phi(t) \rangle
\end{equation}
can be deduced from the Lagrangian
\begin{equation}
\label{qlagrangian}
\mathsf{L}(\Phi, \overset{\bullet}{\Phi}, N) = \langle \Phi(t) | \left( i  \frac{\partial}{\partial t} - {N}  \hat{\mathrm{C}}\right)  |\Phi(t) \rangle \,,
\end{equation}
via the minimization of the respective action with respect to $ |\Phi(t) \rangle$. The quantum counterpart of the classical constraint $\mathrm{C} = 0$ can be obtained as follows:
\begin{equation}
\label{constraint}
- \frac{\partial \mathsf{L}}{\partial {N}} =  \langle \Phi(t) |  \hat{\mathrm{C}}  |\Phi(t) \rangle =0\,.
\end{equation}
The commonly used Dirac method of imposing constraints, $ \hat{\mathrm{C}}  |\Phi(t) \rangle =0$ implies (\ref{constraint}) but the reciprocal does not hold in general. This means that a state $ |\Phi(t) \rangle$ satisfying (\ref{constraint}) does not necessarily lie in the kernel of the operator $ \hat{\mathrm{C}} $.\\
Inspired by Klauder \cite{klauder2012}, we assume that $ |\Phi(t) \rangle $ reads
\begin{equation}
\label{coupledstate}
\left.\begin{array}{c}
|\Phi(t) \rangle = U(Q,q_0) \left( | \tilde{\psi}_{q(t), p(t)} \rangle \otimes | \phi^{{\rm (anis)}} (t) \rangle \right) \\
| \phi^{{\rm (anis)}} (t) \rangle = \sum_n c_n(t) |e_n^{{\rm (anis)}}(q_0) \rangle \,,
\end{array}\right.
\end{equation}
where the different elements are defined as follows: (a) $ | \tilde{\psi}_{q(t), p(t)} \rangle \in \mathcal{H}^{{\rm (iso)}}$ is a $(q,p)$-time-dependent ACS, the fiducial vector $\tilde{\psi}$ being constrained by $\tilde{c}_{-3}=1$ as in the section \ref{sec:sIntemiclass}, (b) $U(Q,q_0)$ is the unitary operator resulting from the substitution $q \mapsto Q$ in the operator defined in (\ref{newunitary}), 
(c) $q_0$ is an arbitrary fixed value of $q$. \\
The unitary operator $U(Q,q_0)$ introduces minimal entanglement (quantum coupling) between the isotropic degree of freedom and anisotropic ones, allowing a more complex quantum behavior than a simple tensor product of states. Replacing  $|\Phi(t) \rangle$ in (\ref{qlagrangian}) by the expression above (\ref{coupledstate}), we obtain the following semiclassical Lagrangian $\mathsf{L}^{{\rm semi}}(q,\dot{q}, p, \dot{p}, \phi^{{\rm (anis)}}, \partial_t \phi^{{\rm (anis)}}, {N})$ (see \cite{QC2016} for more details):
\begin{align}
\label{Newlagrangian}
\nonumber \mathsf{L}^{{\rm semi}}(q,\dot{q}, p, \dot{p},& \phi^{{\rm (anis)}}, \partial_t \phi^{{\rm (anis)}}, {N}) = -q \, \dot{p} +  \langle \phi^{{\rm (anis)}} (t) | i \frac{\partial}{\partial t} | \phi^{{\rm (anis)}} (t) \rangle \\
 - {N} \mathrm{C}_s^{{\rm (iso)}}&(q,p) +{N} \langle \phi^{{\rm (anis)}} | \hat{\mathrm{C}}^{{\rm (anis)}}_s(q,p) | \phi^{{\rm (anis)}}  \rangle
\end{align}
To avoid introducing new unessential constants, we neglect in the sequel the dressing effects of semiclassical formula (functions of $Q$) given in the section \ref{sec:sIntemiclass}. In this case the real function $\mathrm{C}_s^{{\rm (iso)}}(q,p)$ and the operator $\hat{\mathrm{C}}^{{\rm (anis)}}_s(q,p)$ read:
\begin{equation}
\label{semiconstr}
\left.\begin{array}{c}
\mathrm{C}_s^{{\rm (iso)}}(q,p) =  \dfrac{9}{4} p^2+ \tilde{W}(q)\,, \quad \tilde{W}(q) = \dfrac{\tilde{\frak{K}}}{q^2} + W(q) \,,\\
\hat{\mathrm{C}}^{{\rm (anis)}}_s(q,p) = -\dfrac{9}{2}  p\, \hat{\mathsf{A}}(q) +\dfrac{9}{4}  \hat{\mathsf{A}}(q)^2 +\sum_n E_n(q)  | e_n^{{\rm (anis)}} (q_0) \rangle \langle e_n^{{\rm (anis)}} (q_0) |\,,
\end{array}\right.
\end{equation}
where $\hat{\mathsf{A}}(q)$ is the self-adjoint operator defined in (\ref{Afield}).

\subsection{Dynamical equations, Adiabatic and Non-adiabatic approximations}
\begin{figure}[htb]
\sidecaption
\includegraphics[width=0.45\textwidth]{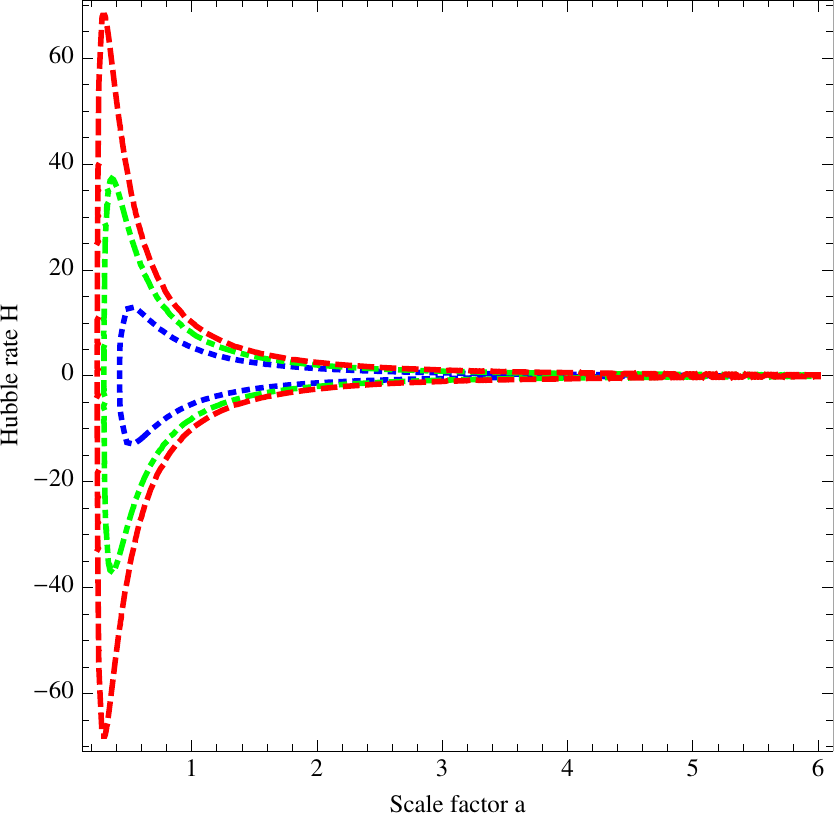} 
\caption{Adiabatic framework: plot of different trajectories (i.e. different values of $Z$) in the plane $(a = q^{2/3}, \texttt{H})$. The classical singularity is replaced by a quantum bounce. (Source: \cite{QC2015b})}
\label{figBIXadiab}
\end{figure}

\begin{figure}[htb]
\sidecaption
\includegraphics[width=0.45\textwidth]{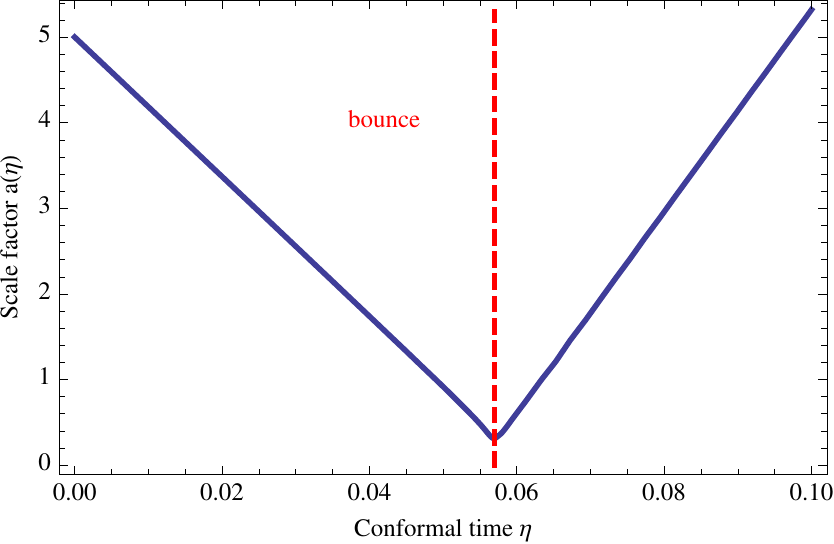}
\hspace{1cm}
\includegraphics[width=0.45\textwidth]{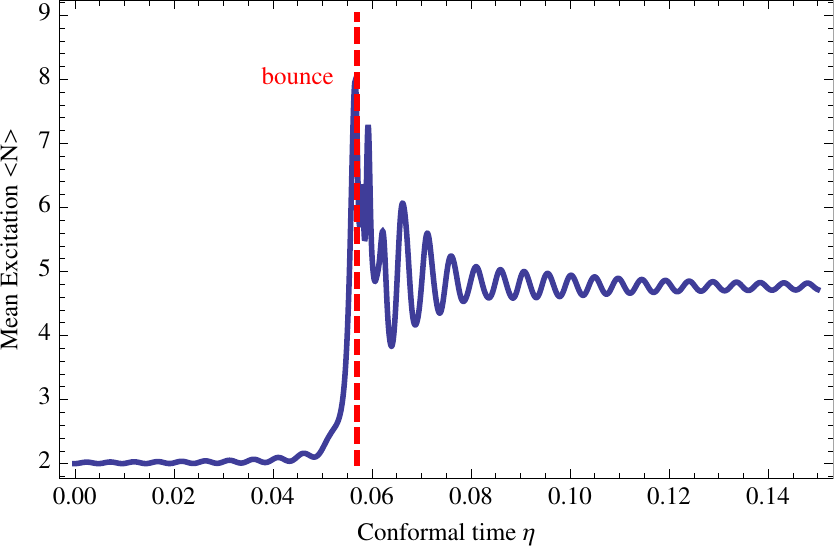} 
\caption{Non-adiabatic framework: on the left panel plot of the scale factor $a=q^{2/3}$ as a function of time during a bounce, on the right panel plot of the degree of excitations through the same bounce. (Source: \cite{QC2016})}
\label{figBIXvib}
\end{figure}

From (\ref{Newlagrangian}) and (\ref{semiconstr}) we deduce the complete set of dynamical equations including the action of the isotropic variable on the anisotropic ones and the backaction of the anisotropic variables on the isotropic one:
\begin{equation}
\label{b9dynamical}
\left.\begin{array}{c}
\dot{q} = {N} \dfrac{\partial}{\partial p}\left( \mathrm{C}_s^{{\rm (iso)}}(q,p) - \langle \phi^{{\rm (anis)}} | \hat{\mathrm{C}}^{{\rm (anis)}}_s(q,p) | \phi^{{\rm (anis)}}  \rangle \right)\\
\dot{p} = - {N} \dfrac{\partial}{\partial q}\left( \mathrm{C}_s^{{\rm (iso)}}(q,p) - \langle \phi^{{\rm (anis)}} | \hat{\mathrm{C}}^{{\rm (anis)}}_s(q,p) | \phi^{{\rm (anis)}}  \rangle \right)\\
i \, \dfrac{\partial}{\partial t}  | \phi^{{\rm (anis)}}  \rangle = - {N} \hat{\mathrm{C}}^{{\rm (anis)}}_s(q,p) | \phi^{{\rm (anis)}}  \rangle
\end{array}\right.
\end{equation}
The classical constraint $\mathrm{C} = 0$ is given in this framework by the semiclassical formula
\begin{equation}
\label{NewQc}
- \frac{ \partial \mathsf{L}^{{\rm semi}}}{\partial {N}} = \mathrm{C}_s^{{\rm (iso)}}(q,p) - \langle \phi^{{\rm (anis)}} | \hat{\mathrm{C}}^{{\rm (anis)}}_s(q,p) | \phi^{{\rm (anis)}}  \rangle = 0
\end{equation}
The Hubble rate $\texttt{H}$ from (\ref{b9dynamical}) reads
\begin{equation}
\texttt{H} = \dfrac{2}{3 {N} } \dfrac{\dot{q}}{q} = \dfrac{3}{q}  \left( p - \langle \phi^{{\rm (anis)}} |\hat{\mathsf{A}}(q)  | \phi^{{\rm (anis)}}  \rangle \right) \,.
\end{equation}
Therefore we obtain from (\ref{NewQc}) the modified Friedman equation
\begin{equation}
\left.\begin{array}{c}
\dfrac{1}{4}\texttt{H} ^2+\dfrac{9}{4 q^2} \sigma_A(q)^2 + \dfrac{\tilde{W}(q)}{q^2}  - \sum_n \dfrac{E_n(q)}{q^2} \left| \langle  \phi^{{\rm (anis)}} | e_n^{{\rm (anis)}}(q_0) \rangle \right|^2  = 0\,,\\
\sigma_A(q)^2 = \langle \phi^{{\rm (anis)}} |\hat{\mathsf{A}}(q)^2  | \phi^{{\rm (anis)}}  \rangle-\left( \langle \phi^{{\rm (anis)}} |\hat{\mathsf{A}}(q)  | \phi^{{\rm (anis)}}  \rangle \right)^2 
\end{array}\right.
\end{equation}
where $\texttt{H} $, $q$ and $ | \phi^{{\rm (anis)}}  \rangle$ are implicitly time-dependent.\\
Since the dynamical system (\ref{b9dynamical}) does not admit explicit analytical solutions, two kinds of approximations can be investigated.
\begin{itemize}
\item[(a)] \textbf{Adiabatic framework} \cite{QC2015a, QC2015b} A detailed analysis of (\ref{b9dynamical}) shows that the operator $\hat{\mathsf{A}}(q)$ is responsible of non-adiabatic effects \cite{QC2016}, i.e. the dynamical coupling between the isotropic variable and the anisotropic ones. Therefore a first approximation consists in neglecting $\hat{\mathsf{A}}(q)$ in the equations (\ref{b9dynamical}). In that case the system becomes separable admitting the following solutions. The Friedman equation reduces to
\begin{equation}
\dfrac{1}{4}\texttt{H} ^2+ \dfrac{\tilde{W}(q)}{q^2} - \dfrac{E_Z(q)}{q^2} = 0\,,
\end{equation} 
where $Z$ is a fixed value of the quantum number $n$, while the state $ | \phi^{{\rm (anis)}} (t)  \rangle$ evolves as
\begin{equation}
 | \phi^{{\rm (anis)}} (t) \rangle = \exp \left(i \int_0^t {N}(\tau) E_Z(q_\tau) {\rm d}\tau \right)  |e_Z^{{\rm (anis)}}(q_0) \rangle \,.
\end{equation}
Only one quantum level $Z$ of the anisotropic Hamiltonian is involved in the dynamics and the eigenenergy $E_Z(q)$ follows adiabatically the change of $q(t)$ during evolution. This corresponds to the Born-Oppenheimer approximation in Molecular Quantum Physics. Thanks to the repulsive part $\propto q^{-2}$ of the potential $\tilde{W}(q)$ and the limit $\lim_{q \to 0} q^2 \, E_Z(q) = 0$, the repulsive effect is always dominant near $q=0$ and the classical singularity is cured. It is replaced by a quantum bounce (see Fig. \ref{figBIXadiab}).

\item[(b)] \textbf{Non-adiabatic (vibronic-like) framework} \cite{QC2016} If we take into account the coupling due to $\hat{\mathsf{A}}(q)$, we allow possible excitations and decays of anisotropic states during evolution. The system cannot be solved analytically anymore and only numerical simulations are available. We assume the state  $ | \phi^{{\rm (anis)}} (t) \rangle$ to be a finite sum $ | \phi^{{\rm (anis)}} (t) \rangle = \sum_n c_n(t) |e_n^{{\rm (anis)}}(q_0) \rangle$,  the functions $c_n(t)$ being numerically calculated. This corresponds to the vibronic framework in Molecular Quantum Physics. This procedure is presented in \cite{QC2016} where we used an harmonic approximation of the potential $V_{IX}(\beta_+,\beta_-)$ near its minimum. This approximation allows to obtain analytical formula for the eigenenergies $E_n(q)$, the eigenvectors $| e_n^{{\rm (anis)}}(q) \rangle$ and the operator $\hat{\mathsf{A}}(q)$. We show in \cite{QC2016} that even if the adiabatic approximation is broken (excitations and decays of anisotropy levels are allowed), the classical singularity is still replaced by a quantum bounce (see Fig. \ref{figBIXvib}).
\end{itemize}

\section{Quantum anisotropy}\label{QA}
In this section we are going to analyze properties of the anisotropic part of the Mixmaster potential (\ref{eqB9pot}). It has three ``open'' ${\sf C}_{3v}$ symmetry directions that can be seen as  three deep ``canyons'', increasingly narrow until  their respective wall edges close up at the infinity whereas their respective bottoms tend to zero (see Fig. \ref{figBIXpot}). The potential $V_{IX}$ is asymptotically  confining except for three directions in which $V_{IX} \to 0$:
\begin{align}\nonumber
&\textrm{(i)} ~~~\beta_-=0, ~~\beta_+ \to \infty~,\\
&\textrm{(ii)} ~~ \beta_+= \beta_-/\sqrt{3}, ~~\beta_- \to -\infty~,\\ \nonumber
&\textrm{(iii)}~\beta_+= -\beta_-/\sqrt{3},~~\beta_- \to \infty~.
\end{align}
It is bounded from below and reaches
its absolute minimum value at
$\beta_\pm=0$, where $V_{IX}=0$. 
Those canyons are problematic at the analytical level, but do not originate a continuous spectrum \cite{QC2017}. One may think of regularizing the potential itself, by applying an integral quantization scheme, specifically the Weyl-Heisenberg one, specific to the full plane symmetry. We expect this procedure to smooth out the potential, specially problematic escape canyons, which can give contribution to non-discrete spectrum of the quantum model.

\subsection{Regularized potential}
For each canonical pair $(\beta_{\pm}, p_{\pm})$ we choose  separable Gaussian weights 
\begin{equation}
\label{smoothPi}
\Pi(\beta_{\pm}, p_{\pm}) = e^{-\frac{\beta_{\pm}^2}{2\sigma_{\pm}^2}}\, e^{-\frac{p_{\pm}^2}{2\tau_{\pm}^2}}\, . \end{equation}
This  yields  manageable formulae with familiar probabilistic content. The ``limit'' Weyl-Wigner case holds as  the widths $\sigma_{\pm}$ and $\tau_{\pm}$ are infinite (Weyl-Wigner is singular in this respect!). Integral Gaussian  quantization yields the quantized form of the potential \eqref{eqB9pot}:
\begin{align}
{A}_{V_{IX}(\beta_+,\beta_{-})}&=\frac13\left(2D_+^4 D_-^{12} e^{4\beta_+}\cosh 4\sqrt{3}\beta_--4
D_+ D_-^3 e^{-2\beta_+}\cosh 2\sqrt{3}\beta_-\right.
\nonumber\\ &\left.+D_+^{16} e^{-8\beta_+}-2D_+^4 e^{4\beta_+}\right) +1,\label{qpot9}
\end{align}
where we have denoted for simplicity:
\begin{equation}
 D_+:=  e^{\frac{2}{\sigma^2_+} },\ \ \ D_-:=  e^{\frac{2}{\sigma^2_-} }.
\end{equation}
The original Bianchi IX potential $V_{IX}(\beta_+,\beta_{-})$ is recovered for $D_+=1=D_-$, thus for weights $\sigma_+,\ \sigma_-\rightarrow \infty$.

\begin{figure}[!ht]
\sidecaption
\includegraphics[width=0.45\textwidth]{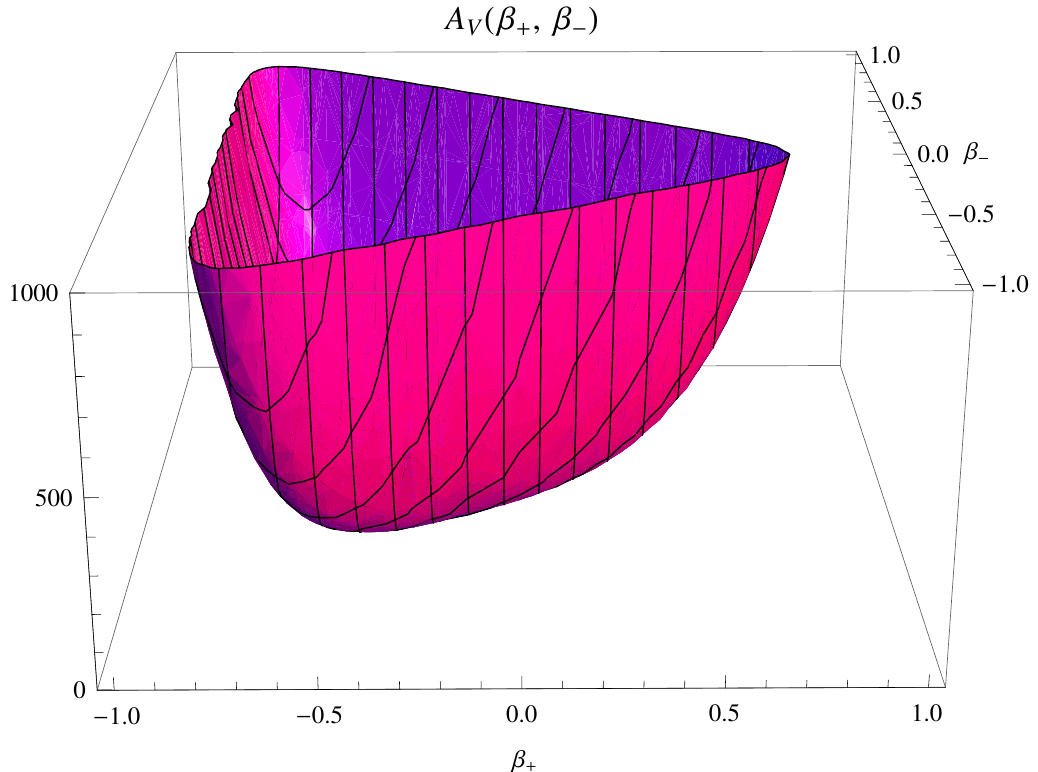} 
\caption{The plot of the regularized Bianchi IX potential near its minimum, for sample values $D_+=1.1$, $D_-=1.4$  (source \cite{intb9}).}
\label{figregpot1}
\end{figure}

Figure \ref{figregpot1} shows the form of the potential \eqref{qpot9} for sample values of $D_+$ and $D_-$. The original escape canyons became regularized and the whole potential is now fully confining. However it has become anisotropic in the variables $\beta_+$ and $\beta_-$ and its minimum  is shifted from the $(0,0)$ position. Imposing either no shift  or isotropy condition yields:
\begin{equation}\label{shiftcon}
 D_+=D_-.
\end{equation}

The full quantized Bianchi IX potential, with implemented condition \eqref{shiftcon} reads as:
\begin{align}\label{isopot}
 {A}_{V_{IX}(\beta_+,\beta_{-})}=\frac13&\left(2D_+^{16}e^{4\beta_+}\cosh 4\sqrt{3}\beta_- -4 D_+^{4}e^{-2\beta_+}\cosh 2\sqrt{3}\beta_-\right.\\&+D_+^{16}e^{-8\beta_+}
 \left.-2D_+^{4}e^{4\beta_+}\right)+1.
\end{align}
The form of this potential is shown in Fig. \ref{figregpot1}. Direct verification shows  it is invariant with respect to  rotations by $2\pi/3$ and $4\pi/3$, thus the $\mathsf{C}_{3v}$ symmetry is preserved and the original isotropy is this way recovered.
\begin{figure}[!ht]
\sidecaption
\includegraphics[width=0.45\textwidth]{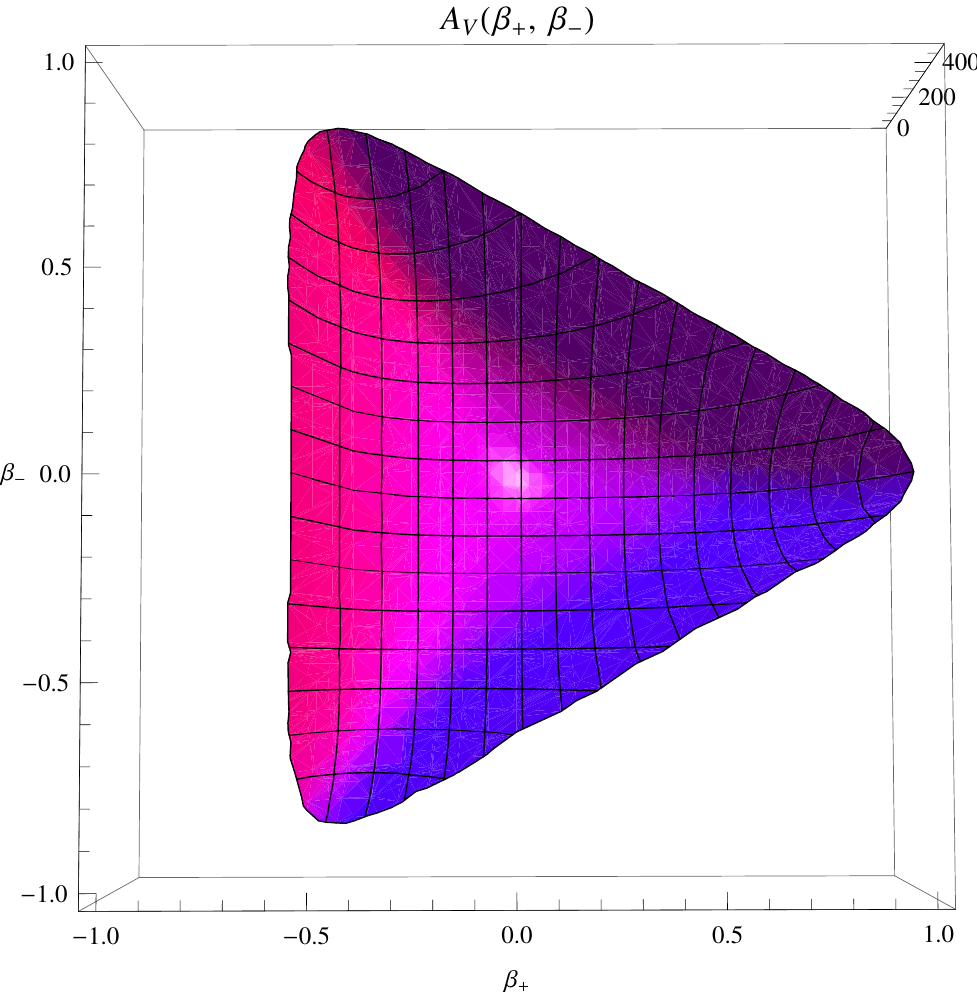} 
\caption{The plot of the regularized Bianchi IX potential near its minimum (source \cite{intb9}).}
\label{figregpot1}
\end{figure}

Further work on the properties of the regularized spectrum  is in progress. Prelimimary results show that the main part of the potential, up to the leading orders, might lead to the integrable dynamic system \cite{intb9}.

\section{Time issue}\label{TI}
The basic feature of the canonical formalism of general relativity is the appearance of a Hamiltonian constraint. In the context of the finite-dimensional Bianchi models studied herein the Hamiltonian constraint is given by a unique term (i.e. the scalar constraint) that is multiplied by a non-vanishing and otherwise arbitrary lapse function $N$. The inclusion of a perfect fluid combined with a particular choice of $N$ makes the constraint linear with respect to the fluid momentum, $p_T$. This enables us to solve the constraint by removing $T$ and $p_T$ from the initial phase space, making $T$ the internal time variable and identifying the non-vanishing Hamiltonian that generates the physical dynamics with respect to $T$. Next, the reduced formalism is quantized (see Sec. \ref{sec3}).

Note that the choice of $T$ for the internal time variable is not unique and there are infinitely many other equally good choices. This property is sometimes called the multiple choice problem and it is discussed e.g. in \cite{Ku}. The free choice of internal time variable constitutes a symmetry of the canonical formalism of gravitational systems. Ultimately, one wants to learn the meaning of this new symmetry which is absent in usual, non-gravitational systems. But first, one needs to find out what kind of differences are induced in the respective descriptions of quantum dynamics derived with different choices of internal time. As we will see, the semiclassical framework based the affine coherent states is useful for this investigation. The following presentation is based on a series of papers devoted to this problem \cite{M1,M2,M3,M4}.

\subsection{Extension to canonical transformations}
The usual way in which one deals with a Hamiltonian constraint is to bring the constraint to the form that is linear with respect to some momentum, say $p_T$:
\begin{equation}
C=p_T+H(T,q,p),
\end{equation}
where $(T,p_T)$ and $(q,p)$ are canonical pairs (see \cite{Ish,Ku} for general discussions). Then, the reduced phase space based on $T$ is given by the canonical pair $(q,p)$ in which the dynamics is generated by $H(T,q,p)$ and occurs in the internal time $T$. This structure is often encoded in the so called contact form which is defined in the contact manifold $(T,q,p)\in\mathcal{C}=\mathbf{R}^3$:
\begin{equation}
\omega_{\mathcal{C}}=\ud q\ud p-\ud T\ud H
\end{equation}
The usual symmetry of the canonical formalism is given by canonical transformations. They are (often time-dependent) transformations of the canonical variables,
\begin{equation}
(q,p)\times T\mapsto (\bar{q},\bar{p}):~~\omega_{\mathcal{C}}=\ud \bar{q}\ud \bar{p}-\ud T\ud \bar{H},
\end{equation}
such that the form of the contact form $\omega_{\mathcal{C}}$ is preserved. Note that the internal time $T$ is preserved too. However, in order to incorporate the freedom in choosing one's internal time into this formalism, one extends the usual symmetry to the so called pseudo-canonical transformations \cite{M1}:
\begin{equation}
(q,p,T)\mapsto (\bar{q},\bar{p},\bar{T}):~~\omega_{\mathcal{C}}=\ud \bar{q}\ud \bar{p}-\ud \bar{T}\ud \bar{H},
\end{equation}
which include the internal time transformations $T\mapsto\bar{T}(q,p,T)$ as well. The group of pseudo-canonical transformations comprises the group of canonical transformations as a normal subgroup. A distinguished subgroup of pseudo-canonical transformations is given by such transformations that preserve the formal expressions for constants of motion \cite{M3}. Namely, if $C(q,p,T)$ is a constant of motion, that is,
\begin{equation}
\partial_TC(q,p,T)-\{C(q,p,T),H(q,p)\}=0,
\end{equation}
then $\bar{C}(\bar{q},\bar{p},\bar{T})=C(\bar{q},\bar{p},\bar{T})$ represents the same constant of motion expressed in terms of another set of contact coordinates and hence,
\begin{equation}
\partial_{\bar{T}}C(\bar{q},\bar{p},\bar{T})-\{C(\bar{q},\bar{p},\bar{T}),H(\bar{q},\bar{p})\}=0.
\end{equation}
(Notice that the Hamiltonian itself is a constant of motions and its form is preserved too.) Let us call this subgroup the `special pseudo-canonical transformations'. They play an important role in defining a certain quantization which is unique for all choices of internal time variable. 

\subsection{Quantization of reduced formalisms}
For a fixed canonical formalism expressed in terms of two sets of canonical variables that are related by a canonical transformation, quantization may sometimes lead to two unitarily equivalent quantum descriptions. In other words, the classical canonical symmetry may be lifted to the quantum level. This, however, is not possible for any two reduced formalisms related by a pseudo-canonical transformation that includes a non-trivial change of the internal time variable. 

A closer look at pseudo-canonical transformations shows that they in general do not preserve the canonical structure. More precisely, any two Poisson brackets, which are based on different internal times, differ for dynamical observables but are the same for conserved observables, i.e. constants of motion. In the context of Hamiltonian constraint systems, the latter are called Dirac observables.

Quantization of reduced formalisms based on different internal times leads to different quantum theories. Nevertheless, it is possible for all the respective quantum descriptions to be seen as an expression of a unique underlying quantum realm (see \cite{M4}). Specifically, since the Dirac observables admit unique Poisson commutation relations they may be given a unique quantum representation on a fixed Hilbert space irrespectively of the choice of internal time. It can be easily shown that the choice of the quantum representation of Dirac observables also fixes the quantization of dynamical observables. However, since the latter lack unique Poisson commutation relations, for a fixed quantum representation of Dirac observables, they are given different quantum operators for different choices of internal time variable (it can be shown with the help of the special pseudo-canonical transformations introduced above, see \cite{M3}). 

As a result of the above quantization prescription, any non-dynamical characterization of a given quantum state is unambiguous for all internal time variables. On the other hand, any dynamical characterization of a given quantum state depends on the internal time employed in the quantum description. Note that the ambiguity concerns only the dynamical interpretation of state vectors in a fixed Hilbert space rather than the quantum dynamics itself as it is generated by a quantized Dirac observable and thus must be unique for all internal time variables. The ambiguity in the physical interpretation of a unique quantum dynamics of a quantum gravitational system is illustrated with a cosmological example below.

\subsection{Semiclassical portraits from different reduced formalisms}
The semiclassical framework based on the affine coherent states is an excellent tool for demonstrating the extent of ambiguity in the quantum dynamics described in different internal times. The idea of the comparison procedure is to confine the unique quantum dynamics to a unique family of the affine coherent states and obtain a semiclassical dynamics in the reduced phase space based on different internal clocks. Some differences in the semiclassical description will appear as a result of different interpretation of any coherent state with respect to different internal time variables.

Given a quantum Hamiltonian, $A_H$, the semiclassical portrait in $(q,p)$ follows from minimization of the following action:
\begin{equation}\label{semA}
\mathbf{S}_{sem}(q,\dot{q},p,\dot{p})=\int \langle q,p|i\frac{\partial}{\partial T}-A_H|q,p\rangle \ud t,
\end{equation}
where $|q,p\rangle$ denotes a specific family of the affine coherent states. As already discussed, the Hamiltonian is a Dirac's observable and is given the same quantum operator $A_H$ in a fixed Hilbert space for all choices of internal time variable. Also, the family of coherent states is unique. Thus, the semiclassical dynamics is unambiguously given by the action (\ref{semA}) in all internal time variables except for that the dynamical interpretation of this dynamics depends on the specific choice of internal time. 

Specifically, the dynamical content of any coherent state $|q,p\rangle$ is provided by the expectation values of the momentum and position operators,
\begin{equation}
q=\langle q,p |\hat{Q}|q,p\rangle,~~p=\langle q,p |\hat{P}|q,p\rangle.
\end{equation}
However, the physical interpretation of the operators $\hat{Q}$ and $\hat{P}$ depends on the employed internal time variable \cite{M2}. Therefore, the semiclassical portraits look only {\it formally} the same in all internal time variables. In order to see any dissimilarities one needs to relate the {\it physical} meanings of the variables featuring in the respective descriptions of the dynamics. 

\begin{figure}[!t]
\includegraphics[width=0.45\textwidth]{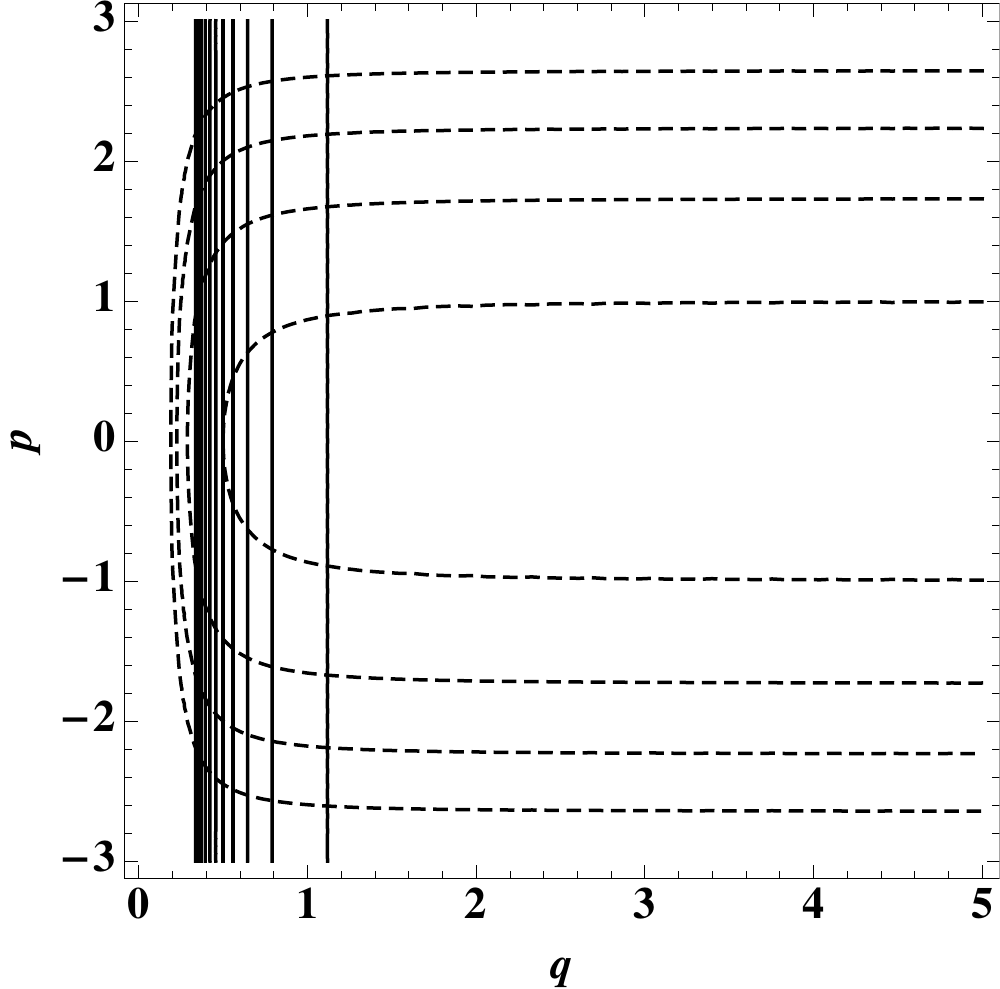}
\hspace{1cm}
\includegraphics[width=0.45\textwidth]{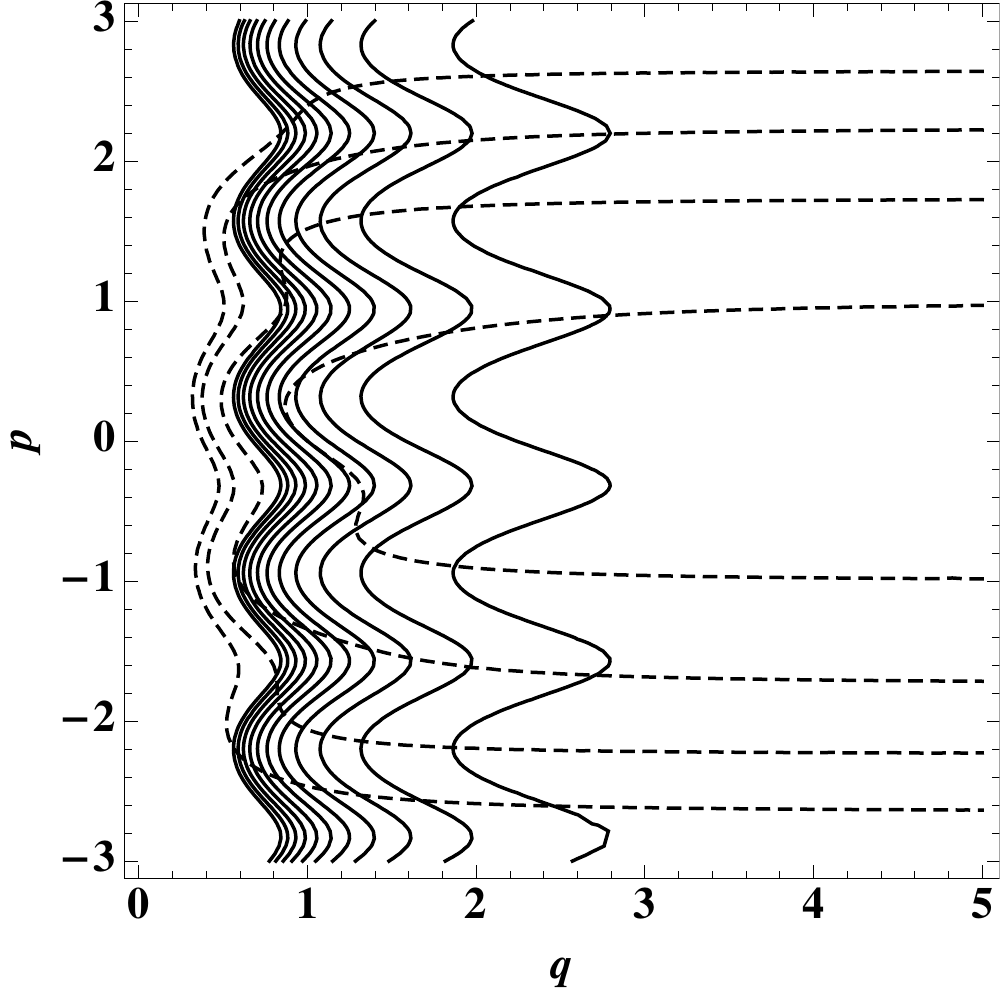}
\caption{The above semiclassical trajectories are generated in different internal time variables defined by some non-trivial delay functions $D(q,p)$. They clearly differ from the original semiclassical trajectories depicted in the right panel of Fig. \ref{fig0}. The vertical lines are the equipotential lines of the repulsive quantum-induced potential which produces the bounces. The plots are taken from \cite{M3}.}
\label{fig1}
\end{figure}

As a concrete example, let us examine the dynamics of a free particle on the half-line $q>0$ generated by the Hamiltonian $\mathrm{H}=p^2$. The following pseudo-canonical transformation,
\begin{equation}\label{freePps}
\bar{T}=T+D(q,p),~~\bar{p}=p,~~\bar{q}=q+2pD(q,p),
\end{equation}
preserves (under some mild assumptions) the range of basic variables, $\bar{q}>0$ and $\bar{p}\in\mathbf{R}$, and the form Hamiltonian, $\mathrm{H}=\bar{p}^2$, which generates the same dynamics with respect to the new internal time $\bar{T}$. The delay function $D(q,p)$ is free (except for some mild restrictions) and encodes the redefinition of the internal time. Eq. (\ref{freePps}) sets a coordinate (i.e., {\it physical}) relation between two reduced formalisms and needs to be applied to the respective semiclassical portraits in order to determine the extent of dissimilarities between the interpretations of the dynamics in different internal time variables. The result of the comparison made for two delay functions is depicted in Fig. \ref{fig1}.

In \cite{M4} (``Internal clock formulation of quantum mechanics") we show that  the quantum formalism based on ambiguous internal time variables allows only for a limited number of physical predictions can be consistently drawn from respective quantum models. Nevertheless, a consistent interpretation of that formalism is possible and it includes some dynamical predictions. In particular, quantum models of singularity resolution are meaningful and can be consistently used for modeling the non-singular evolution of the Universe.

\section{Conclusions}\label{C}
In this short contribution we attempted to show broad applications of the affine coherent states in the analysis of quantum gravitational models. We investigated the fate of classical singularities at the quantum level and the problem of ambiguous internal time variable on which the description of evolution of quantum gravitational models relies. 

The affine coherent states employed for quantization turn out successful in resolving gravitational singularities. In particular, as we have showed, they resolve the oscillatory big-bang/big-crunch singularity which is argued to play a pivotal role in a generic space-like singularity of general relativity. The affine coherent states can be also used to establish a semiclassical framework which allows for analysing the essential features of quantum dynamics. When combined with standard molecular physics approximations, this framework is well-suited for investigating the quantum Mixmaster universe. Furthermore, we have showed that the Weyl-Heisenberg coherent states, if used for quantization of the anisotropic variables, result in regularization of the anisotropy potential and smoothing the problematic ``escape canyons".

The semiclassical framework based on affine CS combined with pseudo-canonical transformations of the internal time variable is a new method for investigating the nature of quantum evolution in gravity. The obtained results clarify some unusual features of quantum mechanics of gravitational systems.

\begin{acknowledgement}
We wish to thank the organizers of the conference ``Coherent States and their Applications: A Contemporary Panorama" for their invitation and for a friendly and stimulating atmosphere during the meeting.
\end{acknowledgement}

\end{document}